\documentclass[pdflatex,sn-mathphys-num]{sn-jnl}

\usepackage{graphicx}%
\usepackage{multirow}%
\usepackage{amsmath,amssymb,amsfonts}%
\usepackage{amsthm}%
\usepackage{mathrsfs}%
\usepackage[title]{appendix}%
\usepackage{xcolor}%
\usepackage{textcomp}%
\usepackage{manyfoot}%
\usepackage{booktabs}%
\usepackage{algorithm}%
\usepackage{algorithmicx}%
\usepackage{algpseudocode}%
\usepackage{listings}%

\usepackage{booktabs}
\usepackage{longtable}
\usepackage{siunitx}
\usepackage{hyperref}

\theoremstyle{thmstyleone}%
\theoremstyle{thmstyletwo}%

\theoremstyle{thmstylethree}%

\begin{document}

\title[Article Title]{
Thermodiffusion in Aqueous Alkali Halide Solutions from Ambient to Supercooled Conditions: Ion-Specific, Structural, and Mass Effects}

\author{\fnm{Guansen} \sur{Zhao}}\email{g.zhao21@imperial.ac.uk}
\author{\fnm{Fernando} \sur{Bresme}}\email{f.bresme@imperial.ac.uk}

\affil{\orgdiv{Department of Chemistry}, \orgname{Imperial College London}, \orgaddress{\street{Molecular Sciences Research Hub}, \city{London}, \postcode{W12 0BZ}, \country{United Kingdom}}}

\abstract{Thermodiffusion in aqueous electrolyte solutions exhibits complex dependencies on temperature, concentration, and salt composition, yet its microscopic origins remain incompletely understood. Here, we employ non-equilibrium molecular dynamics (NEMD) simulations to investigate thermal transport and thermodiffusion in aqueous alkali halide solutions over the temperature range 240--300~K at concentrations of 1~m and 4~m. Building on previous studies of NaCl and LiCl, we extend the analysis to systems containing K$^+$ and I$^-$ ions to assess ion-specific effects. Across all systems studied, the thermal conductivity decreases upon cooling and is generally reduced at higher salt concentration.
The Soret coefficient generally increases with temperature, shifting the solutions from thermophilic behavior at low temperature toward more thermophobic behavior at high temperature. Clear ion-dependent trends are observed, with Na$^+$ and K$^+$ salts generally showing stronger thermophobic responses than Li$^+$ salts, especially in iodide solutions. We estimate that the shift in the inversion temperatures of the iodide salts relative to experiment corresponds to a small local offset of the effective heat of transport,
 4-5 kJ\,mol$^{-1}$, showing that small changes in hydration thermodynamics or heat--mass coupling can strongly affect the sign change of the Soret coefficient. Structural analyses indicate that lower temperatures and lower concentrations favor more tetrahedrally ordered, LDL-like water environments, which are associated with enhanced thermophilicity. Analysis of inversion temperatures and mass effects further suggests that the heat of transport contains both structural and kinetic contributions. These findings provide molecular-level insight into the interplay between hydration structure, ionic mass, and thermodiffusive transport in aqueous electrolytes.}


\keywords{Molecular dynamics, Non-equilibrium thermodynamics, Thermodiffusion, Electrolytes, Thermal conductivity, Thermal transport}

\maketitle

\section{Introduction}\label{sec:intro}
   
The study of electrolyte solutions under non-isothermal conditions began with the seminal work of Ludwig and Soret~\cite{Ludwig1856,Soret1879}, who showed that thermal gradients induce salt concentration gradients. The extent of this separation is quantified by the Soret coefficient, which describes the tendency of solutes to migrate toward either hot regions (thermophilic behavior) or cold regions (thermophobic behavior). Within the framework of linear non-equilibrium thermodynamics~\cite{deGrootMazur1984,kjelstrup_non-equilibrium_2008}, the Soret coefficient is defined as \( s_T = D_T / D_{12} \), where \(D_T\) and \(D_{12}\) are the thermodiffusion and interdiffusion coefficients, respectively. A negative Soret coefficient (\(s_T < 0\)) corresponds to thermophilic behavior, whereas a positive value (\(s_T > 0\)) indicates thermophobic behavior. In liquid mixtures, the magnitude of the Soret coefficient is typically on the order of $10^{-3}$--$10^{-2}\,\mathrm{K^{-1}}$ for organic and aqueous solutions~\cite{kita_sign_2004,koniger_measurement_2009,cabrera_measurement_2013,romer_alkali_2013}, while values approaching $10^{-1}\,\mathrm{K^{-1}}$ have been reported for biomolecular systems~\cite{piazza_thermophoresis_2008,vigolo_thermophoresis_2010,duhr_why_2006,baaske_extreme_2007,wiegand_thermal_2004}.

Thermodiffusion in aqueous electrolytes has attracted significant interest because temperature gradients can induce the migration of ions, which can be exploited to drive mass separation processes, including the desalination of seawater-like solutions~\cite{xu_thermodiffusive_2024} and renewable energy applications~\cite{perez_de_luco_mass_2025}. Recent experimental and simulation studies have shown that the Soret coefficient of aqueous electrolyte solutions exhibits complex dependencies on temperature, concentration, and ion identity. In many alkali halide solutions, the Soret coefficient can change sign as a function of temperature or concentration, indicating a transition between thermophilic and thermophobic regimes~\cite{romer_alkali_2013,di_lecce_thermal_2018,mohanakumar_overlapping_2022,bresme_thermal_2024}. In addition, a pronounced minimum in the Soret coefficient as a function of salt concentration has been observed in chloride and iodide systems, typically at intermediate molalities~\cite{colombani_thermal_1999,di_lecce_role_2017,rudani_analyzing_2025}. Recent studies of ammonium chloride salts have also highlighted the connection between hydrophilicity and thermodiffusion~\cite{rudani_hydrophilicity_2026}.

Despite the advances outlined above, our ability to predict thermodiffusion in ionic solutions from theory remains limited. By contrast, computer simulations of simple nonionic mixtures modeled with the Lennard-Jones potential have provided important insight into the microscopic factors that govern the Soret coefficient. Differences in particle mass, size, and intermolecular interaction strength have all been shown to modify the thermodiffusive response of binary mixtures~\cite{hafskjold_molecular_1993,kincaid_thermal_1994,reith_nature_2000,artola_microscopic_2007}. In particular, the interaction asymmetry and mass ratio between the components can strongly affect both the magnitude and the sign of the Soret coefficient by introducing an asymmetry in energy transport between species~\cite{reith_nature_2000}. Extending these ideas to ionic solutions, previous simulation studies confirmed the existence of a minimum in the Soret coefficient of LiCl solutions~\cite{di_lecce_role_2017,dilecce_computationalapproach_2017}, which had earlier been identified experimentally~\cite{colombani_thermal_1999}. These studies further showed that disruption of the tetrahedral hydration structure of lithium suppresses this minimum. However, the role of ionic mass in determining the Soret coefficient has received far less attention. To our knowledge, only one simulation study of LiCl has examined the effect of ion mass by varying the cation and anion masses independently~\cite{di_lecce_role_2017}. That work found that ionic mass has a relatively weak influence on the Soret coefficient compared with ion--water interactions and hydration-shell structure, particularly for the strongly hydrated Li$^+$ ion. More broadly, mass effects have rarely been addressed in studies of thermodiffusion in electrolyte solutions.

Most thermodiffusion studies to date have focused on systems under standard temperature and pressure conditions. However, electrolytes also give rise to important colligative effects, such as the depression of water's freezing point. This effect has important consequences because it enables investigations at low temperatures that are difficult to access in pure water~\cite{kanno_homogeneous_1977,conde_molecular_2018}, where homogeneous ice nucleation limits experimental studies. Exploring aqueous systems in this temperature range is particularly important because water, unlike simple liquids, exhibits anomalous structural and thermodynamic properties that become especially pronounced under supercooled conditions~\cite{Debenedetti_metastable_1996,Eisenberg_structure_2005}. These anomalies are often interpreted in terms of enhanced density fluctuations as the liquid approaches a hypothesized second critical point. Within the liquid--liquid phase transition (LLPT) framework, this behavior is associated with a first-order transition between locally ordered low-density liquid (LDL) and more disordered high-density liquid (HDL) environments, terminating at a liquid--liquid critical point~\cite{poole_phase_1992,gallo_water_2016}. We note that the existence of a liquid-liquid critical point in water has been verified very recently~\cite{you_experimental_2026}. 

The dual structural character of water appears to play a central role in its anomalous thermal transport properties, particularly its thermal conductivity, as suggested by recent studies~\cite{gittus_unravelling_2025,bresme_communication_2014,zhao_thermal_2024}. In the deeply supercooled regime, these transport anomalies have been linked to the competition between local HDL and LDL structures and to the associated anomalous response functions~\cite{bresme_communication_2014,zhao_thermal_2024}. The existence of these two competing structural motifs, favored at high and low temperatures, respectively, may also provide insight into the microscopic variables underlying the complex thermodiffusive behavior of aqueous ionic solutions. Indeed, previous studies have discussed the thermodiffusion of aqueous solutions in relation to the hydrogen-bond network of water~\cite{niether_thermophoresis_2019}. Very recent work on supercooled aqueous solutions further showed that the Soret coefficient exhibits a nonmonotonic temperature dependence in NaCl and LiCl solutions, with this behavior being closely linked to the increasing dominance of LDL-like structures at low temperatures~\cite{zhao_thermal_2025}. These findings highlight the strong coupling between thermodiffusive transport and the microscopic structure of water, thereby motivating further investigation across a broader range of ions and thermodynamic states.

In this work, we employ non-equilibrium molecular dynamics (NEMD) simulations to investigate thermal transport and thermodiffusion in aqueous electrolyte solutions. Building on previous studies of NaCl and LiCl~\cite{zhao_thermal_2025}, we extend the analysis to systems containing K$^+$ and I$^-$ ions, enabling a systematic assessment of ion-specific effects. The simulations are performed over the temperature range 240--300~K at both dilute (1~m) and higher (4~m) concentrations. We further characterize the structural properties of water, with particular emphasis on the hydrogen-bond network, to elucidate its correlation with thermophilicity in these solutions. Finally, we examine the influence of ion mass on the thermodiffusive response in order to assess the relative importance of mass effects and ion--water interactions. To isolate mass effects, we keep the ion--water interactions unchanged while varying the masses of the cations and anions. We find that mass effects are more significant than previously recognized, inducing a transition from thermophilic to thermophobic behavior as the salt mass increases and thereby highlighting the relevance of purely kinetic contributions.


\section{Methods}\label{sec:method}

Water was modeled using the rigid, nonpolarizable TIP4P/2005 model~\cite{abascal_general_2005}, which is known to provide an accurate representation of water over a wide range of thermodynamic conditions, including ambient liquid water, supercooled states, and several ice polymorphs~\cite{vega_simulating_2011,vega_what_2008}. Ion--ion and ion--water interactions were described using the Madrid-2019 force field~\cite{zeron_force_2019,blazquez_madrid-2019_2022}, a parameterization designed for use with TIP4P/2005 that incorporates scaled ionic charges to reproduce the thermodynamic and structural properties of electrolyte solutions. This combination of force fields has successfully reproduced the thermal transport behavior of aqueous alkali salt solutions under ambient conditions~\cite{bresme_thermal_2024} and has also been widely applied to modeling NaCl and LiCl solutions over a broad temperature range, from under-ambient to deeply supercooled conditions~\cite{perin_phase_2023,sedano_isothermal_2024,zhao_thermal_2025}.

\begin{figure}[h]
\centering
\includegraphics[width=0.9\textwidth]{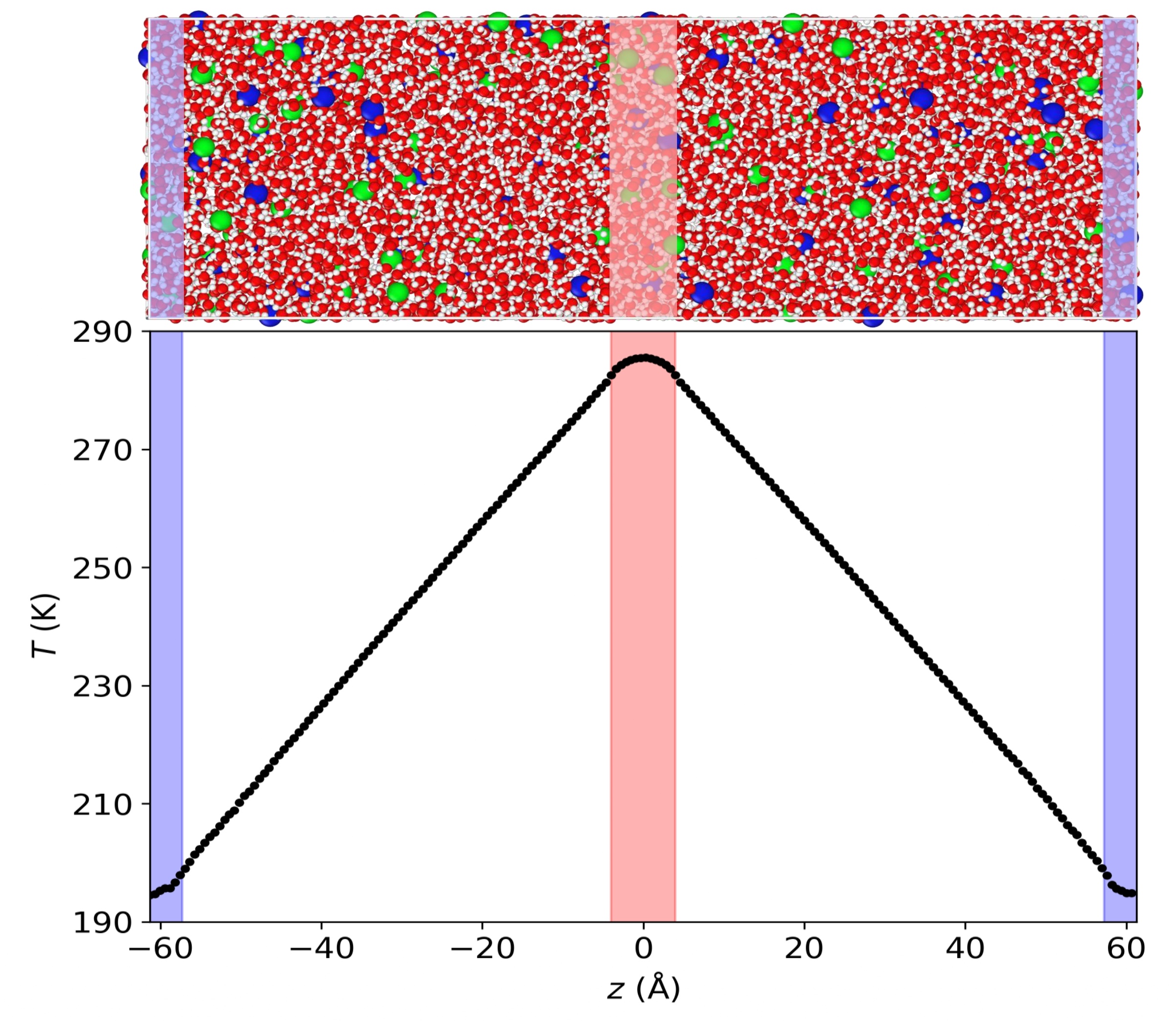}
\caption{(Top) Snapshot of the simulation cell used in a representative NEMD simulation of a 1~m LiI aqueous solution, visualized using OVITO~\cite{stukowski_visualization_2010}. The configuration corresponds to an average temperature of 241~K, a density of 1.089~g\,cm$^{-3}$, and an average pressure of 15.7~bar during the simulation run. Oxygen and hydrogen atoms in water are shown as red and white spheres, respectively, while Li$^{+}$ and I$^{-}$ ions are depicted in blue and green. The thermostatting regions are highlighted in red and blue, corresponding to the hot slab at the center of the simulation box and the cold slabs near the edges, respectively. (Bottom) Temperature profile along the $z$-direction for the imposed temperature difference.}
\label{fig:sim_box}
\end{figure}

Following the simulation protocol established in our previous work~\cite{zhao_thermal_2025}, boundary-driven non-equilibrium molecular dynamics (NEMD) simulations were performed using LAMMPS (version 29 Aug 2024)~\cite{plimpton_fast_1995} to investigate thermal transport and the associated coupling effects in aqueous alkali halide solutions. Temperature gradients were imposed by defining hot (\(T_\text{hot}\)) and cold (\(T_\text{cold}\)) reservoirs using Langevin thermostats~\cite{schneider_molecular-dynamics_1978,brunger_stochastic_1984}, located at the center and near the edges of the simulation box, respectively (see Figure~\ref{fig:sim_box}). The thermostats were applied at every 1~fs timestep, with a coupling constant of 500~fs. To prevent drift, the total center-of-mass momentum of the system was removed at each timestep. Each thermostatting slab had a thickness of 8~\text{\AA}, and the simulation cell was elongated along the direction of the imposed gradient, with box-length ratios of \(L_x:L_y:L_z = 1:1:3\). The transverse box length, \(L_x\), varied from 40.90 to 42.95~\text{\AA} because of changes in the system density.

Long-range electrostatic interactions were treated using the particle--particle particle--mesh (PPPM) method~\cite{Hockney1989}, with a relative accuracy of \(10^{-5}\). The same approach was used to account for long-range dispersion interactions~\cite{isele-holder_development_2012,isele-holder_reconsidering_2013}, with accuracy thresholds of \(10^{-4}\)~kcal/(mol\,\AA) in real space and \(2\times10^{-4}\)~kcal/(mol\,\AA) in reciprocal space.

Typical simulation cells contained 6663 water molecules together with either 120 or 480 ion pairs, corresponding to salt concentrations of 1 and 4~m, respectively. Before applying the thermal gradients, each system was pre-equilibrated for 5~ns in the isothermal-isobaric (NPT) ensemble at 1~bar and at a temperature equal to the average of the target values of $T_\text{hot}$ and $T_\text{cold}$, using a Nos\'e-Hoover thermostat and barostat with relaxation times of 0.1 and 1~ps, respectively~\cite{nose_molecular_1984,hoover_canonical_1985}. The subsequent NEMD simulations were then performed under constant volume conditions.

Based on our previous simulations of supercooled TIP4P/2005 water at 1~bar, which showed that the system reaches a steady state within approximately 10--20~ns at temperatures above 230~K~\cite{zhao_thermal_2024}, all simulations in the present work were run for at least 100~ns. For the systems investigated here, with average temperatures of 240 and 300~K, the initial 20~ns was discarded as further equilibration. During the production runs, local temperature and density profiles were obtained by averaging over 200 spatial bins in the \(z\)-direction, corresponding to bin widths ranging from \(0.613\) to \(0.644~\text{\AA}\). Statistical precision was further improved by performing 10 independent replicas with different thermostat random seeds,  resulting in a total sampling time of 0.8~$\mathrm{\mu}$s. Trajectories from the final 10~ns of each replica were saved every 100 timesteps, yielding a total of $10^{6}$ configurations for the calculation of tetrahedral order parameters and electrostatic potential profiles.

The thermal conductivity, $\lambda(z)$, was evaluated using Fourier’s law,
\begin{equation}
\lambda(z) = - \frac{J_q}{\nabla T(z)} ,
\label{eqn:fourier}
\end{equation}
where $\nabla T(z)$ denotes the local temperature gradient at position \(z\) along the direction of the heat flux $J_q$. This gradient was obtained from a linear fit to the steady-state temperature profile (see the bottom panel of Figure~\ref{fig:sim_box}), performed over a temperature window of $\pm 5$~K around the target temperature. The heat flux was computed from the continuity equation,
\begin{equation}
J_q = \frac{\dot{Q}}{2A}
\label{eqn:heat-flux}
\end{equation}
where $\dot{Q}$ is the time derivative of the energy transferred between the hot and cold reservoirs, and $A$ is the cross-sectional area of the simulation cell perpendicular to the direction of heat flux. The factor of two accounts for the presence of two oppositely directed heat fluxes in the simulation box (see Figure~\ref{fig:sim_box}). Energy conservation was carefully monitored and was satisfied within numerical accuracy, with equal magnitudes of energy exchanged at the hot and cold thermostats (see Figure~\ref{fig:energy_conserv} in Appendix~\ref{secA:energy_conserv}).

The Soret coefficient $s_T$ was determined from the steady-state salt mole fraction profiles under conditions of vanishing solute flux ($J_1 = 0$),
\begin{equation}
   s_T(T)= -\frac{1}{x_1(T)\,\left (1-x_1(T)\right )}\left ( \frac{\nabla x_1(T)}{\nabla T(T)} \right )_{J_1 =0}
   = -\frac{1}{b(T)} \left ( \frac{db(T)}{dT} \right )_{J_1 =0} ,
\label{eqn:Soret}
\end{equation}
where $x_1(T)$ denotes the salt mole fraction at the local temperature $T$, and $b(T)$ is the corresponding local molality. The gradients in Eq.~\eqref{eqn:Soret} were obtained from linear fits to the local temperature and salt mole fraction profiles evaluated within a \(\pm 5~\mathrm{K}\) window around the target temperature.

Structural changes in the electrolyte solutions were investigated by computing the tetrahedral order parameter $\xi$, defined as~\cite{errington_relationship_2001}
\begin{equation}
\xi = 1 - \frac{3}{8} \sum_{j=1}^{3} \sum_{k=j+1}^{4} \left( \cos \psi_{jk} + \frac{1}{3} \right)^2 ,
\label{eqn:order_param}
\end{equation}
where $\psi_{jk}$ denotes the angle formed by the vectors connecting a reference water molecule to its four nearest neighboring oxygen atoms. According to Eq.~\eqref{eqn:order_param}, \(\xi = 1\) corresponds to perfect tetrahedral coordination (\( \psi_{jk} = 109.47^\circ \)), whereas \(\xi \approx 0\) indicates a random, gas-like angular distribution. The fractions of LDL- and HDL-like environments were obtained from the probability distribution of $\xi$ following the procedure described in Ref.~\cite{zhao_thermal_2025}.

To further characterize the structural organization of the electrolyte solutions, oxygen-oxygen radial distribution functions were computed from equilibrium molecular dynamics simulations performed in the isothermal-isobaric (NPT) ensemble using GROMACS (version Jan 2024)~\cite{Gromacs1993,GROMACS2024}. Each system consisted of a cubic simulation box containing 926 water molecules and 17 or 67 ion pairs, corresponding to concentrations of 1 and 4~m, respectively. Production simulations were carried out for a minimum of 100~ns with three independent replicas performed to ensure adequate statistical sampling. A timestep of 1~fs was used. Temperature and pressure were controlled using the Nosé-Hoover thermostat~\cite{nose_molecular_1984,hoover_canonical_1985} and the Parrinello-Rahman barostat~\cite{parrinello_polymorphic_1981}, with coupling time constants of 2 and 4~ps, respectively.

\section{Results}\label{sec:results}

\begin{figure}[ht]
\centering
\includegraphics[width=1\textwidth]{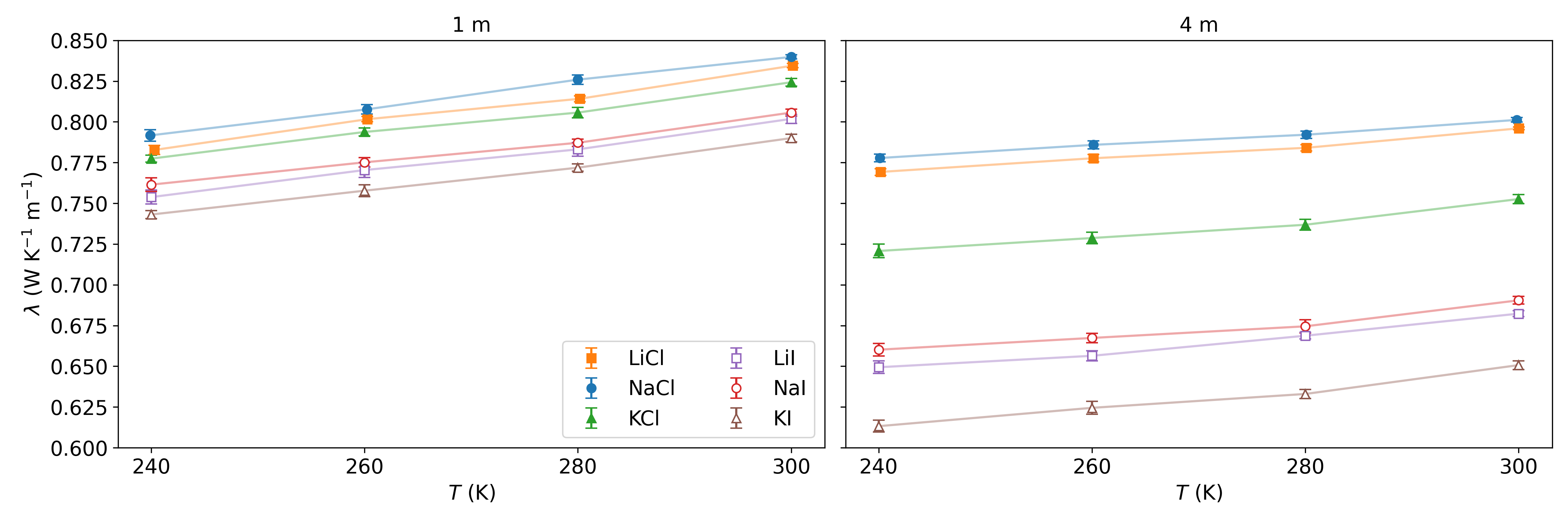}
\caption{Thermal conductivity of aqueous NaCl, LiCl, KCl, NaI, LiI, and KI solutions over 240--300~K at concentrations of 1~m (left) and 4~m (right). Results for NaCl and LiCl are taken from our previous work~\cite{zhao_thermal_2025}. Values are averaged over ten independent NEMD replicas, each analyzed in successive 10~ns trajectory segments, with error bars representing the standard deviation across replicas. Lines are included as visual guides to the eye.}
\label{fig:tc}
\end{figure}

Figure~\ref{fig:tc} shows the thermal conductivity of aqueous electrolyte solutions as a function of temperature over the 240--300~K range. As the temperature decreases from ambient conditions, the thermal conductivity of all solutions decreases. These results therefore follow the temperature dependence of thermal transport in water, consistent with the previously reported reduction in thermal conductivity upon cooling~\cite{biddle_thermal_2013,bresme_communication_2014,zhao_thermal_2024}. Comparison of the 1~m and 4~m data indicates that increasing salt concentration generally reduces the thermal conductivity. At 1~m, differences among the salts reveal a clear anion-dependent trend. Solutions containing chloride ions (NaCl, LiCl, and KCl) exhibit higher thermal conductivities, whereas iodide solutions show markedly lower values.

The larger reduction in thermal conductivity observed for iodide salts follows the well-known experimental trend that salts with higher molar masses produce a greater decrease in solution thermal conductivity~\cite{abdulagatov_thermal_2001}. Previous studies have shown that iodide salts, which contain larger and heavier anions, are more compressible and exhibit a slight decrease in the speed of sound as the concentration increases, resulting in lower thermal conductivity than in chloride salts and a stronger reduction with increasing salt concentration~\cite{bresme_thermal_2024}. Consistent with this trend, the 4~m iodide solutions remain less thermally conductive than the corresponding chloride solutions, and potassium salts exhibit noticeably lower conductivities than the analogous sodium and lithium systems.

\begin{figure}[h]
\centering
\includegraphics[width=1\textwidth]{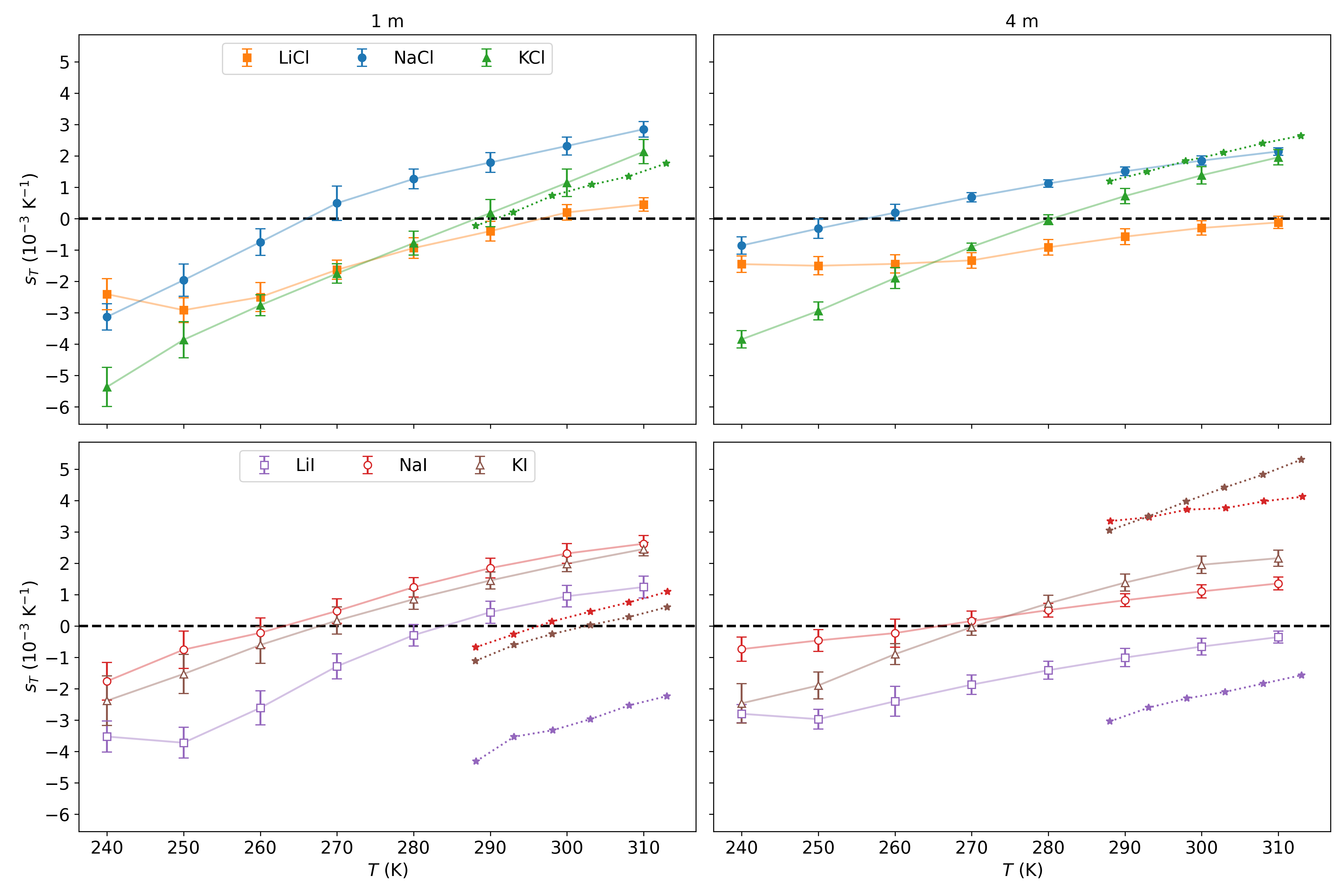}
\caption{Soret coefficients of aqueous electrolyte solutions over the 240--300~K temperature range. Chloride salts (NaCl, LiCl, KCl) are shown in the top row and iodide salts (NaI, LiI, KI) in the bottom row, with 1~m concentrations in the left column and 4~m in the right column. The horizontal dashed black line denotes zero Soret coefficient. Results for NaCl and LiCl are reproduced from our previous work~\cite{zhao_thermal_2025}. Star symbols connected by dotted lines show experimental data: iodide salts (NaI and KI) in the range $\sim$288--313~K from Ref.~\cite{mohanakumar_overlapping_2022}, and KCl from Ref.~\cite{mohanakumar_thermodiffusion_2021}.}
\label{fig:soret}
\end{figure}

In Fig.~\ref{fig:soret}, we show the Soret coefficients as a function of temperature for aqueous electrolyte solutions to examine salt-dependent thermodiffusion. Overall, our results follow previous simulation studies at ambient temperature~\cite{bresme_thermal_2024}, reproducing both the magnitude and the temperature dependence of the Soret effect in aqueous alkali halide solutions. For all systems studied here, the Soret coefficient increases with temperature, leading to a transition from thermophilic behavior at low temperatures to thermophobic behavior at high temperatures. Such thermophilic-thermophobic transitions have been previously reported in both experimental and simulation studies of aqueous electrolyte solutions using a variety of force fields~\cite{gaeta_nonisothermal_1982,romer_alkali_2013,di_lecce_thermal_2018,mohanakumar_overlapping_2022,rudani_analyzing_2025,sugaya_thermal_2006}.

A more detailed comparison with experimental measurements~\cite{mohanakumar_overlapping_2022} is provided for the iodide salts shown in the bottom panels of Fig.~\ref{fig:soret}. At 1~m, the simulations reproduce the experimental ordering of the Soret coefficients, with NaI exhibiting the strongest thermophobic response, followed closely by KI, while LiI displays significantly smaller values. All three systems show a monotonic decrease in the Soret coefficient with decreasing temperature above approximately 250~K, consistent with the experimental trends. Quantitatively, however, the thermophilic-thermophobic inversion observed experimentally near 300~K for NaI and KI occurs at temperatures roughly 30~K higher than those predicted by the simulations, and the simulated Soret coefficients are systematically larger in magnitude than the experimental values for all three salts. At 4~m, the agreement with experiments becomes less quantitative: the simulations overestimate the Soret coefficient for LiI, whereas NaI and KI are slightly underestimated relative to the measurements. Nevertheless, the overall ion-dependent trend remains consistent with the experiment. Taken together, these comparisons indicate that the Madrid-2019 ion parameters combined with the TIP4P/2005 water model capture the main experimental trends in the thermodiffusion of iodide salts, while exhibiting systematic deviations in the absolute magnitude of the Soret coefficients.

In Fig.~\ref{fig:shifted_T} in Appendix~\ref{secB:rescale}, we present an alternative comparison with experimental data for 1~m NaI, KI, and KCl solutions, for which experimental inversion temperatures are available, by shifting the simulation temperatures to match \(T_{inv}\), defined by \(s_T(T_{inv}) = 0\). The required temperature shift is strongly anion-dependent: while the iodide salts exhibit a substantial offset (\(\sim 30\)~K), KCl shows only a minor shift (\(\sim 2\)~K). After this rescaling, the agreement between simulation and experiment becomes nearly quantitative in the three cases, indicating that the models employed here accurately capture the temperature dependence of the Soret coefficient near the inversion temperature.


The inversion point has particular significance because the Soret coefficient is directly linked to the heat of transport ~\cite{eastman_thermodynamics_1926,eastman_theory_1928, deGrootMazur1984,shukla_new_1998,dougherty_theory_1955}, which measures the energy transferred across a reference plane per mole of transported component, in excess of the convective enthalpy associated with that transport. The heats of transport of solute and solvent are related to each other via the Gibbs-Duhem equation (see Refs.~\cite{shukla_new_1998,dougherty_theory_1955}):

\begin{equation}
x_1 Q_1^* = -x_2 Q_2^*,
\label{eq:heats}
\end{equation}
\noindent
where  \(x_i\) and \(Q_i^*\) are the mole fraction and heat of transport of the solute (``1'') and solvent (``2''). In the steady state and at relatively low concentrations, \(\sim 1\) molal, the solute provides the main contribution to the heat of transport. 
At 1 molal, the salt mole fraction is small (\(x_1 \approx 0.018\)), so Eq.~\ref{eq:heats} gives
\(Q_2^* = -(x_1/x_2)Q_1^*\), implying that the solvent contribution is only about \(2\%\) of the solute contribution. Hence, the solute heat of transport dominates the
heat-of-transport difference entering \(s_T\). Agar et al. \cite{agar_thermal_1960,agar_single-ion_1989} have discussed the concept of heat of transport in the context of the Soret coefficient of electrolyte solutions,

\begin{equation}
s_T = \frac{Q^*}{ 2 R T^2 \Gamma}
\label{eq:soret-heat}
\end{equation}
where Eq.~\ref{eq:soret-heat} is written in the Hittorf frame of reference, and the solute heat of transport is defined as \(Q^*= Q_+^* +  Q_-^*\) for monovalent ions. The factor of 2 in the denominator takes into account that there is 1 mol of cation and 1 mol of anion per mol of salt, and \(\Gamma \) is the thermodynamic factor,
\begin{equation}
\Gamma
=
1+\left(\frac{\partial \ln \gamma_{\pm}}{\partial \ln m}\right)_{T,P}
\label{eq:thermodynamic-factor}
\end{equation}
where \( \gamma_{\pm} \) is the solute mean activity coefficient and \(m\) is the salt concentration. 


At the inversion temperature, \(s_T(T_{inv})=0\), we can use Eq.~\ref{eq:soret-heat} to get the gradient \(ds_T/dT\),

\begin{equation}
\left.\frac{d s_T}{dT}\right|_{T_{\mathrm{inv}}}
=
\frac{1}{2 R T_{\mathrm{inv}}^{2}\,\Gamma(T_{\mathrm{inv}})}
\left.\frac{d Q^* }{dT}\right|_{T_{\mathrm{inv}}},
\label{eq:gradient-st}    
\end{equation}
which shows that the slope around the inversion point is directly proportional to the temperature derivative of the effective solute heat of transport \(Q^*\). Assuming that the thermodynamic factor \(\Gamma\) remains of order unity and varies little near \(T_{inv}\) at \(1\) molal, the \(\sim 30\) K difference between simulated and experimental inversion temperatures reported above is unlikely to originate primarily from the denominator \(R T^2\Gamma\) in the expression for \(ds_T/dT\). Instead, it points to a shift in the temperature dependence of the heat-of-transport, \( Q^*(T) \). Since \(s_T(T_{\mathrm{inv}})=0\) implies \(Q^*=0\), the higher experimental inversion temperature indicates that the simulation predicts the zero crossing of \( Q^*(T)\) at too low a temperature in the case of the iodide salts. Linearizing around \(T_{\mathrm{inv}}\) gives,
\begin{equation}
\delta Q^* \approx 2 R T_{\mathrm{inv}}^{2}\Gamma
\left.\frac{ds_T}{dT}\right|_{T_{\mathrm{inv}}}
\delta T_{\mathrm{inv}}.
\label{eq:linearising}
\end{equation}
Using \(\left(ds_T/dT\right)_{T_{\mathrm{inv}}}\sim 10^{-4}\,\mathrm{K}^{-2}\), together with \(\Gamma\sim1\) and \(\delta T_{\mathrm{inv}}\approx30\) K, yields an estimated offset \(\delta Q^*\) of approximately \(4 \text{--}5\ \mathrm{kJ\,mol^{-1}}\). This value should be considered an order of magnitude estimate of the mismatch
in the effective heat of transport that controls the inversion temperature, not as an error in the hydration enthalpy. Although the estimated local offset in \(Q^*\) near the inversion temperature is not large \(4\!-\!5\ \mathrm{kJ\,mol^{-1}}\), such a shift can have a significant
large effect because \(Q^*(T)\) crosses zero at \(T_{\mathrm{inv}}\). In this regime,
small changes in the temperature dependence of ion hydration, hydrogen-bond network
reorganization, or kinetic heat--mass coupling could shift the zero crossing of \(Q^*\)
appreciably and thereby produce a significant shift in the inversion temperature.

The inversion-temperature discrepancy, therefore, reflects a small but systematic mismatch in the simulated temperature dependence of the heat of transport \(Q^*\). This effect is most pronounced for the iodide salts, whereas for KCl the discrepancy is minimal and the results match the experiment almost quantitatively. We infer from this analysis that the force field employed here for NaI and KI does not fail primarily in its description of overall hydration thermodynamics, but rather in the more delicate temperature-dependent enthalpy–entropy balance of ion hydration, highlighting the sensitivity of the Soret coefficient to hydration effects. Following our analysis, relatively small changes in \(Q^*\) translate into significant shifts in the inversion temperature. Despite these differences, it is remarkable that the force fields employed here reproduce the main experimental thermodiffusive trends, including the stronger thermophilic character of lithium salts.



Returning to our simulation results, a comparison across different cations reveals clear ion-specific trends in the thermodiffusive response of the solutions. Under most thermodynamic conditions considered here, salts containing Na$^+$ and K$^+$ exhibit stronger thermophobic behavior than those containing Li$^+$. This trend is consistent with the well-established differences in cation hydration strength, which decrease in the order Li$^+>$ Na$^+>$ K$^+$. Lithium ions form particularly strong and structured hydration shells, whereas the larger alkali metal ions interact more weakly with surrounding water molecules due to their lower charge density and size~\cite{mahler_study_2012}. Previous experimental and simulation studies of alkali halide solutions have highlighted the important role of ion hydration and solute--solvent interactions in determining both the magnitude and sign of the Soret coefficient~\cite{mohanakumar_overlapping_2022,romer_alkali_2013,di_lecce_role_2017}. Consistent with this picture, Fig.~\ref{fig:soret} also shows crossings between the LiCl and KCl solutions at both concentrations near 270~K. In particular, lithium salts exhibit a weaker temperature dependence of the Soret coefficient and a less pronounced decrease upon cooling, especially for chloride systems and at higher concentrations, which gives rise to the observed crossings with KCl.

A comparison of the chloride and iodide salts further highlights the influence of the anion on thermodiffusion. Although the overall magnitudes of the Soret coefficients are similar for the two halides, the temperature at which the thermophilic-to-thermophobic inversion occurs varies less among the iodide salts, particularly for NaI and KI. By contrast, Li$^+$ solutions retain significantly higher inversion temperatures even in the iodide systems, especially at higher concentrations, a result consistent with experimental observations (see Fig.~\ref{fig:soret}). This behavior may reflect the more diffuse hydration structure of iodide ions, which imposes weaker orientational constraints on nearby water molecules than chloride ions and consequently induces a greater disruption of the hydrogen-bond network~\cite{robertson_molecular_2003}. As a result, the strongly hydrated Li$^+$ ion appears comparatively less sensitive to changes in the anionic environment.

Previous studies have highlighted the important role of the hydrogen-bond network of water in determining thermodiffusive behavior in aqueous solutions. In particular, a more strongly organized hydrogen-bond network has been associated with thermophilic responses in such systems~\cite{niether_thermophoresis_2019,zhao_alkali_2025}.   \citet{sugaya_thermal_2006} investigated aqueous urea solutions, in which urea is known to modify water structure by weakening local water--water interactions. At high urea concentrations, this weakening of the hydrogen-bond network leads to thermophobic behavior, in a manner analogous to the effect of increasing temperature.

\begin{figure}[h]
\centering
\includegraphics[width=1\textwidth]{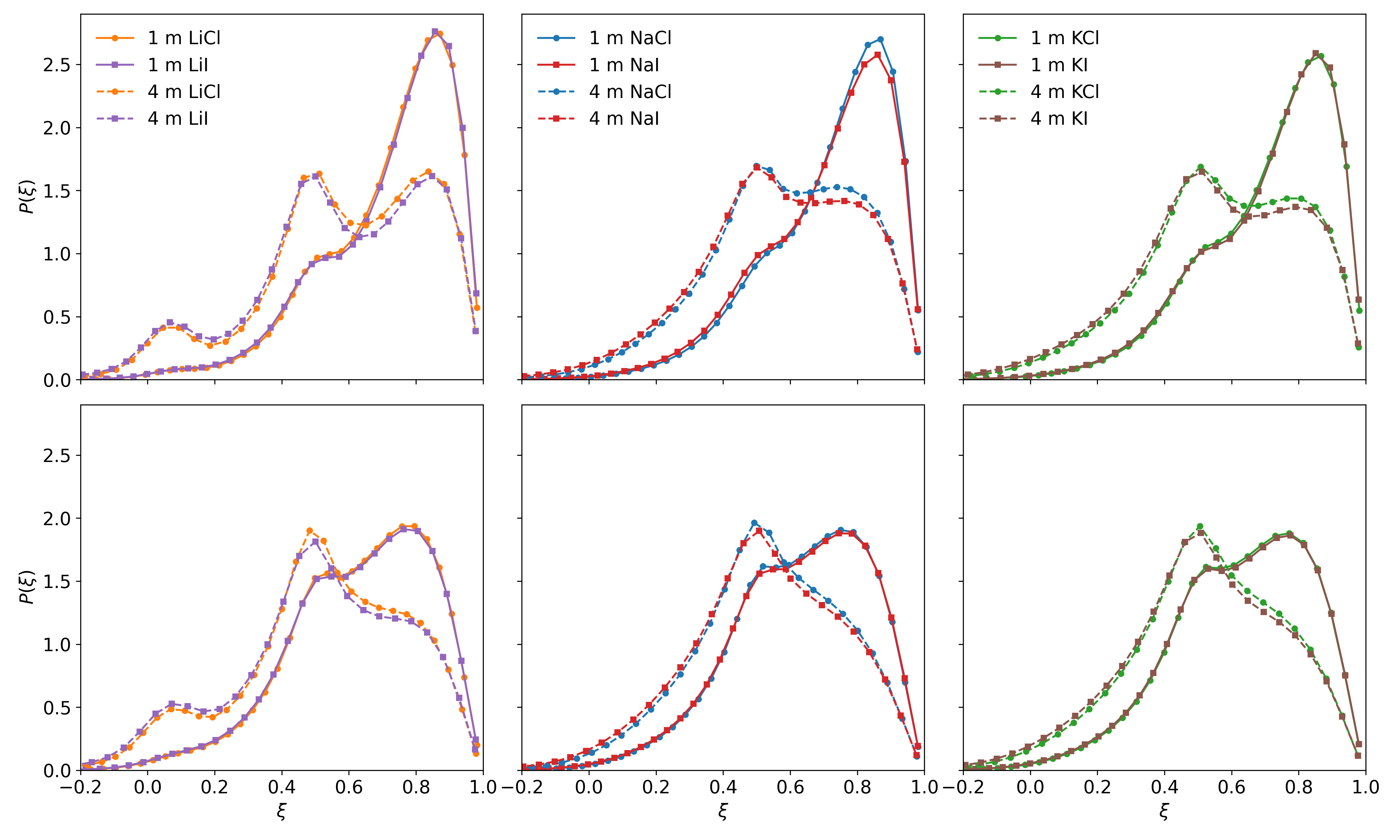}
\caption{Probability density distributions of the tetrahedral order parameter, \(\xi\), for aqueous chloride and iodide salt solutions at fixed cation identity. Results are shown for Li$^{+}$ (left column), Na$^{+}$ (middle column), and K$^{+}$ (right column). The top and bottom rows correspond to temperatures of 240~K and 300~K, respectively. Solid and dashed lines denote concentrations of 1~m and 4~m. Results for NaCl and LiCl were previously reported in our earlier study~\cite{zhao_thermal_2025}.}
\label{fig:tetrahedral}
\end{figure}

In our previous simulations of supercooled NaCl and LiCl solutions, we further identified a correlation between thermophilicity and the populations of two distinct local structural motifs in water~\cite{zhao_thermal_2025}. To investigate whether similar structural signatures arise across the electrolyte solutions considered here, we analyze the distribution of the tetrahedral order parameter \(\xi\). Fig.~\ref{fig:tetrahedral} presents the probability density distributions of \(\xi\) for all systems studied, enabling a direct comparison of cation and anion effects on the local structure of water. The distributions exhibit a bimodal character, commonly observed in supercooled water, that reflects the coexistence of distinct local environments~\cite{errington_relationship_2001,kuo_tetrahedral_2021}. The two dominant peaks correspond to more tetrahedrally ordered configurations associated with low-density-liquid-like (LDL-like) arrangements and to less ordered high-density-liquid-like (HDL-like) environments.

Consistent with our previous work~\cite{zhao_thermal_2025}, lower temperature and lower salt concentration generally lead to a larger population of LDL-like local environments, accompanied by higher averaged tetrahedral order parameters (see Table~\ref{tab:nemd} in Appendix~\ref{secA:table}). These trends indicate a more structured hydrogen bonding network under these conditions. This observation is consistent with the correlation between enhanced tetrahedral organization of the hydrogen-bond network and increased thermophilicity in aqueous electrolyte solutions (see Fig.~\ref{fig:soret}).

Comparing the distributions across different cations reveals that Na$^+$ and K$^+$ solutions exhibit very similar tetrahedral order-parameter distributions, whereas Li$^+$ solutions show a distinct peak at \(\xi < 0.2\). A similar multimodal distribution was previously reported for LiCl solutions~\cite{zhao_thermal_2025} and was attributed to strong ion--water interactions that locally distort the hydrogen-bond network and generate heterogeneous structural motifs. Despite the presence of this highly disordered population, LiI solutions retain a slightly larger average tetrahedral order parameter and a higher fraction of LDL-like structures than the NaI and KI solutions, particularly at the higher concentration and lower temperature of 4~m. This behavior suggests that the small, strongly hydrated Li$^+$ ion tends to preserve more of the overall tetrahedral organization of water outside its first solvation shell. The comparison between chloride and the more diffuse iodide anion is consistent with this picture, as the larger and more weakly hydrated I$^-$ ion induces a slightly greater disruption of the hydrogen-bond network than Cl$^-$, with average reductions of \(\sim\)1--3\% in \(\bar{\xi}\) and 1--5\% in the LDL fraction. 

\begin{figure}[h]
\centering
\includegraphics[width=1\textwidth]{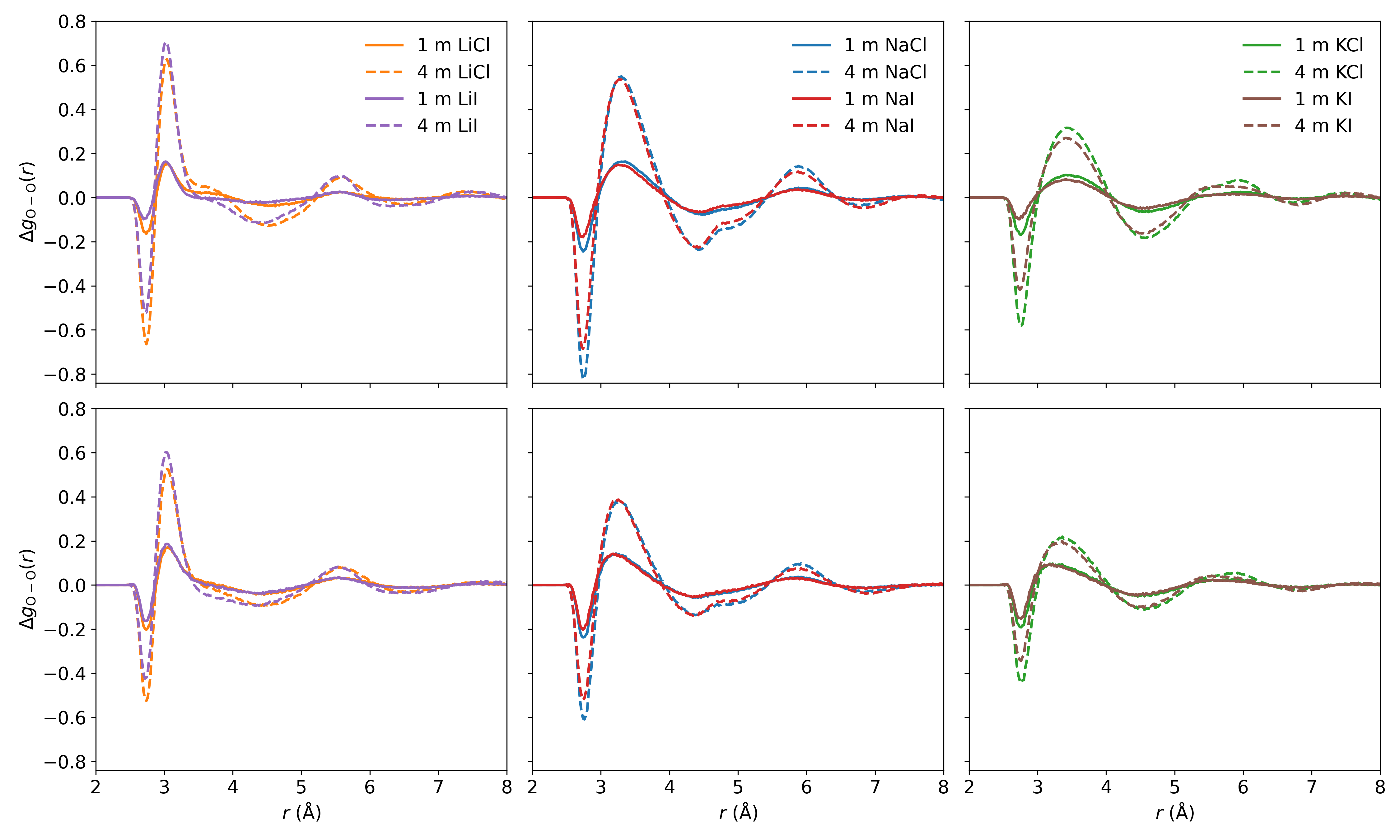}
\caption{Perturbation of the oxygen–oxygen radial distribution function in aqueous salt solutions relative to pure water at the corresponding temperature, defined as \(\Delta g_{OO}(r) = g_{OO,\mathrm{solution}}(r) - g_{OO,\mathrm{pure\ water}}(r)\). Panels are arranged by cation identity, with Li$^{+}$, Na$^{+}$, and K$^{+}$ shown in the left, middle, and right columns, respectively. The upper and lower rows correspond to temperatures of 240~K and 300~K, respectively. Solid curves represent 1~m solutions, while dashed curves correspond to 4~m. The NaCl and LiCl results were obtained by analyzing simulation trajectories from our previous work~\cite{zhao_thermal_2025}. The oxygen--oxygen radial distribution functions of pure water used as reference are provided in Fig.~\ref{fig:water_goo_reference} in Appendix~\ref{secA:OOrdf}.}
\label{fig:O-O_rdf}
\end{figure}

The perturbations of the oxygen-oxygen radial distribution functions in Fig.~\ref{fig:O-O_rdf} further support the structural trends identified from the tetrahedral order parameter analysis. At both temperatures, the addition of salt leads to negative deviations of \(\Delta g_{OO}(r)\) in the \(4\text{--}5\)~\AA\ region, indicating a weakening of the hydrogen-bond network structure~\cite{bresme_thermal_2024,zhao_alkali_2025}. While these changes are relatively small at 1~m, they become clearly more pronounced at 4~m. 
The effect is more pronounced at 240~K (top panels), where the larger magnitude of the negative deviations indicates that ion-induced perturbations more strongly disrupt the hydrogen-bond network. However, the hydrogen-bond organization in the solutions remains stronger at 240~K than at 300~K (see Fig.~\ref{fig:tetrahedral}), indicating that temperature plays a more dominant role in determining the overall network structure than ion-induced perturbations.

Differences between cations are also evident. Li$^{+}$ solutions produce the largest short-range perturbations in the vicinity of the first hydration shell of water (around \(r \approx 3\)~\AA), consistent with the strong and localized hydration structure of Li$^{+}$. By comparison, Na$^{+}$ and K$^{+}$ induce structural perturbations that extend over a wider spatial range, as shown in Fig.~\ref{fig:O-O_rdf}. These trends are reproduced in both chloride and iodide salts, with only minor differences. The longer-range perturbations observed for the sodium and potassium salts are consistent with their stronger thermophobic response relative to the lithium salts.


\begin{figure}[h]
\centering
\includegraphics[width=1\textwidth]{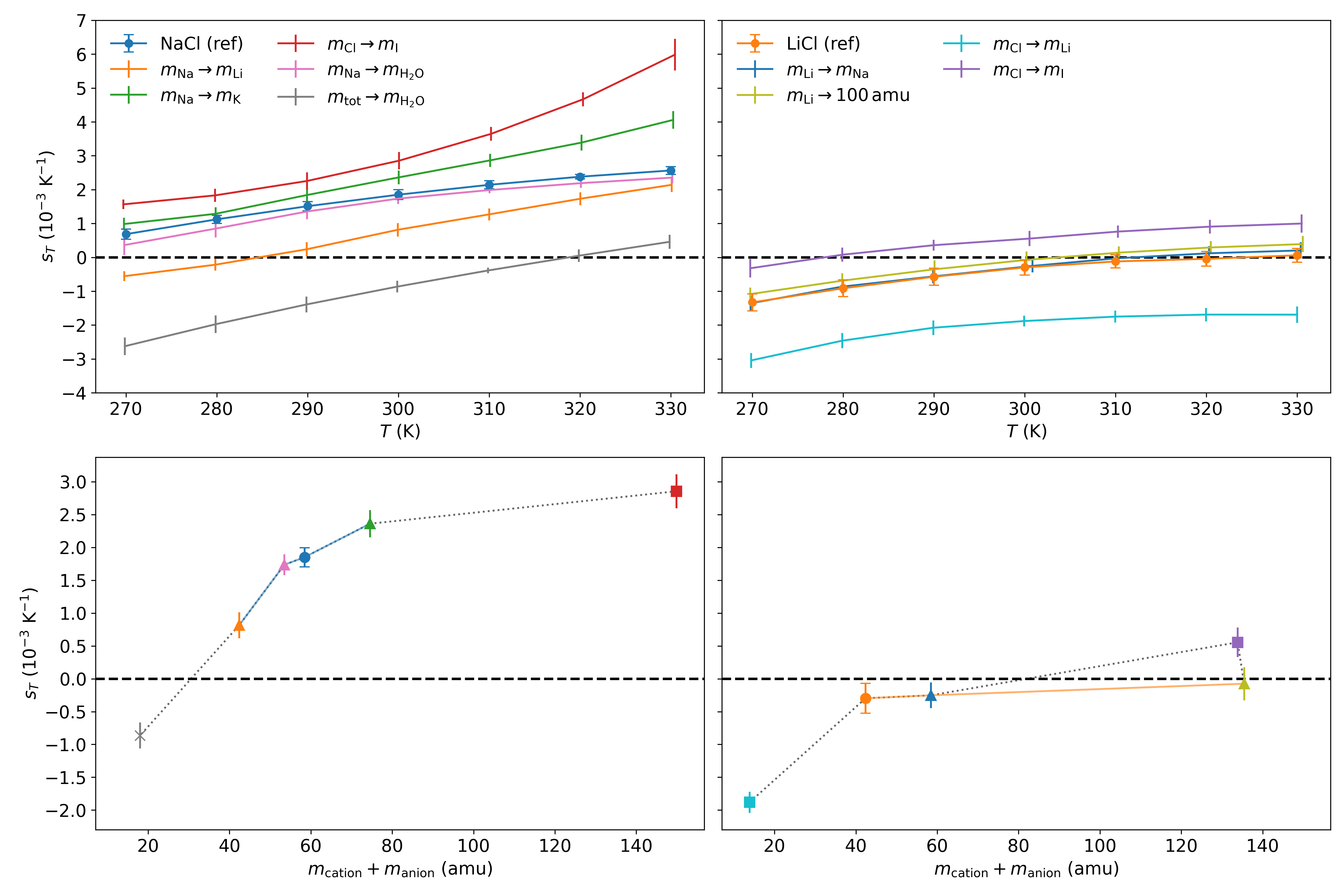}
\caption{Effect of ionic mass on the Soret coefficient \(s_T\) of aqueous electrolyte solutions. The top panels show the temperature dependence of the Soret coefficient between 270 and 330~K for 4~m NaCl (left) and LiCl (right) solutions. These data are taken from our previous work~\cite{zhao_thermal_2025}. Results for the reference systems are shown alongside cases in which the ionic masses were modified. The case labeled \(m_{\mathrm{tot}} \rightarrow m_{\mathrm{H_2O}}\) corresponds to setting the cation and anion masses equal to each other and equal to one half of the water molecular mass. The bottom panels show the Soret coefficient at 300~K as a function of the total ion mass. Circles denote the reference systems, triangles indicate cases where the cation mass is modified, squares correspond to modifications of the anion mass, and crosses represent cases where both ionic masses are changed. The dotted black lines connect all data points, while the solid colored lines connect systems with the same anion mass. The dashed horizontal line indicates \(s_T=0\).}
\label{fig:mass_effect_soret}
\end{figure}

Finally, in Fig.~\ref{fig:mass_effect_soret} we analyze the effect of ionic mass on thermodiffusion. To isolate this contribution, we start from the 4~m NaCl and LiCl reference systems and systematically modify the ionic masses while keeping all interaction parameters unchanged. Because we employ classical (non‑quantum) models, the underlying solution structure, including ion hydration and the hydrogen‑bond network of water, remains unaffected. This allows us to attribute any changes in the Soret coefficient solely to kinetic effects arising from the altered ionic masses.

Our results show that increasing the ionic mass generally shifts the Soret coefficient toward more thermophobic behavior, indicating that heavier ions preferentially accumulate in the cold region. In addition, modifying ionic masses alters the temperature at which the Soret coefficient changes sign, demonstrating that the thermal diffusion coefficient $D_T$ is sensitive to variations in ionic mass.
This behavior is consistent with previous molecular dynamics studies of binary Lennard–Jones (LJ) mixtures, which showed that mass asymmetry alone can influence both the magnitude and sign of the Soret coefficient~\cite{hafskjold_molecular_1993,kincaid_thermal_1994,reith_nature_2000}. In these systems, lighter particles possess larger thermal velocities and transport kinetic energy more efficiently along a temperature gradient, leading to a net migration of heavier species toward the cold region. Correspondingly, tuning the solute-solvent mass ratio can modify the thermal diffusion coefficient and even reverse the sign of the Soret coefficient when the mass ratio crosses unity~\cite{reith_nature_2000}.

The present results indicate that mass-driven shifts in thermodiffusive behavior remain consistent with the kinetic mechanisms identified in simple liquids. Moreover, our results support the idea that the heat of transport $Q^*$ contains a kinetic contribution in addition to the thermodynamic one.  
Because modifying ionic masses leaves hydration thermodynamics unchanged, yet still shifts the Soret coefficient and its inversion temperature, $Q^*$ in ionic solutions must therefore include a mass-dependent kinetic contribution, a conclusion consistent with recent experimental analyses~\cite{rudani_analyzing_2025}. Note, however, that this effect must be system-dependent. Indeed, we showed recently that the existence of minima in the Soret coefficient of non-aqueous liquid mixtures was coincident with a minimum in the thermodynamic factor, and therefore with non-ideal behavior in the mixture~\cite{gittus_microscopic_2023}.

The bottom panels of Fig.~\ref{fig:mass_effect_soret} further illustrate ion-specific effects by correlating the Soret coefficient with the total ionic mass. For the NaCl systems, modifying either the cation or anion mass leads to noticeable changes in the Soret coefficient. For the LiCl systems, however, a different behavior emerges. Modifying the mass of the cation while keeping the anion mass unchanged produces only minor variations in the Soret coefficient, even when the Li$^+$ mass is artificially increased to large values. This weak sensitivity to the cation mass is consistent with previous molecular dynamics simulations of LiCl aqueous solutions employing the SPC/E water model and Dang ion parameters, which reported that ion mass plays a relatively minor role in determining the Soret coefficient~\cite{di_lecce_role_2017}. By comparison, altering the mass of the chloride ion results in a measurable change in the Soret coefficient, indicating that the thermodiffusive response of LiCl solutions is more sensitive to the anion mass than to the cation mass. A possible explanation lies in the strong hydration structure of Li$^+$ in water. Lithium ions are known to form tightly bound hydration complexes with a well-defined coordination number of approximately four, commonly described as $[\mathrm{Li(H_2O)_4}]^+$~\cite{rempe_hydration_2000,zeron_force_2019}. It appears that lithium's strong hydration, {\it i.e.} its particularly strong water–ion interactions, offsets mass effects in this case and makes the thermodiffusive response less sensitive to changes in the bare ionic mass.

\section{Conclusion}\label{sec:conclusion}

In this work, we employed non-equilibrium molecular dynamics (NEMD) simulations to investigate thermal transport and thermodiffusion in aqueous alkali halide solutions over the temperature range 240--300~K at concentrations of 1~m and 4~m. By extending previous studies of NaCl and LiCl to systems containing K$^+$ and I$^-$ ions, we provided a systematic assessment of ion-specific effects on both thermal conductivity and the Soret coefficient across standard conditions and the water supercooled regime.

Our results show that the thermal conductivity decreases monotonically upon cooling across all systems considered and is generally reduced at higher salt concentration. The Soret coefficient exhibits a pronounced temperature dependence, with all solutions undergoing a transition from thermophilic behavior at low temperatures towards thermophobic behavior at higher temperatures. Both the magnitude of the Soret coefficient and the inversion temperature depend strongly on ion type. Salts containing Na$^+$ and K$^+$ generally display stronger thermophobic responses than Li$^+$ systems, while iodide salts exhibit distinct inversion trends relative to their chloride counterparts.

Analysis of the tetrahedral order parameter distributions reveals a correlation between thermodiffusive behavior and the hydrogen‑bond network of water. Lower temperatures and lower salt concentrations promote more tetrahedrally ordered, LDL‑like local environments, which are associated with enhanced thermophilicity. Differences between ions can be rationalized in terms of their hydration structure. In most cases, strongly hydrated Li$^{+}$ salts tend to better preserve the overall hydrogen‑bond organization of water at low temperatures. In contrast, larger and more weakly hydrated cations in iodide salts induce greater disruption of the hydrogen‑bond network and are correspondingly more thermophobic than Li$^{+}$ salts. These results support the view that thermodiffusion in aqueous electrolyte solutions is coupled to ion‑specific modifications of the local water structure, particularly changes in its tetrahedral organization. Thermodynamic conditions in the range 240--300~K, which favor stronger orientational ordering associated with LDL‑like water, therefore yield more thermophilic alkali halide solutions.

The analysis of the inversion temperatures further suggests that the heat of transport governing the sign change of the Soret coefficient is sensitive to both ion type and the underlying hydration thermodynamics. Comparison with experimental data for sodium and potassium iodide salts shows that the simulations capture the overall temperature dependence and ion‑specific trends of the Soret response, but display a systematic shift in the inversion temperature. Because the inversion temperature is defined by \(Q^*(T_{\mathrm{inv}})=0\), we find that even a relatively small local shift in the heat of transport, of order \(4\!-\!5\ \mathrm{kJ\,mol^{-1}}\) for the iodide salts, can shift the Soret inversion temperature appreciably and produce the observed tens of kelvin discrepancy in \(T_{\mathrm{inv}}\), highlighting the sensitivity of thermodiffusion to subtle changes in hydration structure and kinetic heat--mass coupling. This discrepancy is strongly anion-dependent, stronger for Na and Li iodide salts and much smaller for KCl. For the latter we obtain nearly quantitative agreement with the experiment. We therefore conclude that the force fields employed here (TIP4P/2005 and Madrid 2019) capture the essential physics underlying thermodiffusion. The remaining differences relative to the experiment are most likely associated with a delicate temperature‑dependent balance between hydration enthalpy and entropy, rather than with a failure to describe the overall hydration thermodynamics. Our results indicate that analyzing the gradient of the Soret coefficient near the inversion temperature may provide a route to systematically improve the accuracy of existing force fields.

The analysis of mass effects shows that increasing ionic mass generally shifts the Soret coefficient towards more thermophobic behavior, consistent with previously identified kinetic mechanisms in simple liquids. However, the sensitivity to mass depends strongly on ion type. In LiCl solutions, the Soret coefficient is only weakly affected by changes in the lithium mass, suggesting that the strong hydration of Li$^+$ suppresses purely inertial contributions and makes ion--water interactions the dominant factor. These findings indicate that the heat of transport in aqueous electrolytes includes both structural and kinetic contributions.

Overall, this work highlights the interplay between ion-specific hydration, hydrogen-bond network organization, and kinetic effects in determining thermodiffusive transport in aqueous electrolyte solutions. From an experimental perspective, systematic investigations of supercooled ionic solutions may provide a practical route to test these predictions and to examine whether variations in the Soret coefficient remain correlated with changes in hydrogen-bond ordering in the supercooled regime.

\backmatter

\section*{Data Availability Statement}
All data relevant to this study are included in the manuscript and its appendices. 

\bmhead{Acknowledgements}
We acknowledge the ICL RCS High-Performance Computing facility and the UK Materials and Molecular Modelling Hub for computational resources, partially funded by the EPSRC (Grant Nos. EP/P020194/1 and EP/T022213/1).

\begin{appendices}

\section{Energy Conservation Check}\label{secA:energy_conserv}

\begin{figure}[H]
\centering
\includegraphics[width=0.7\textwidth]{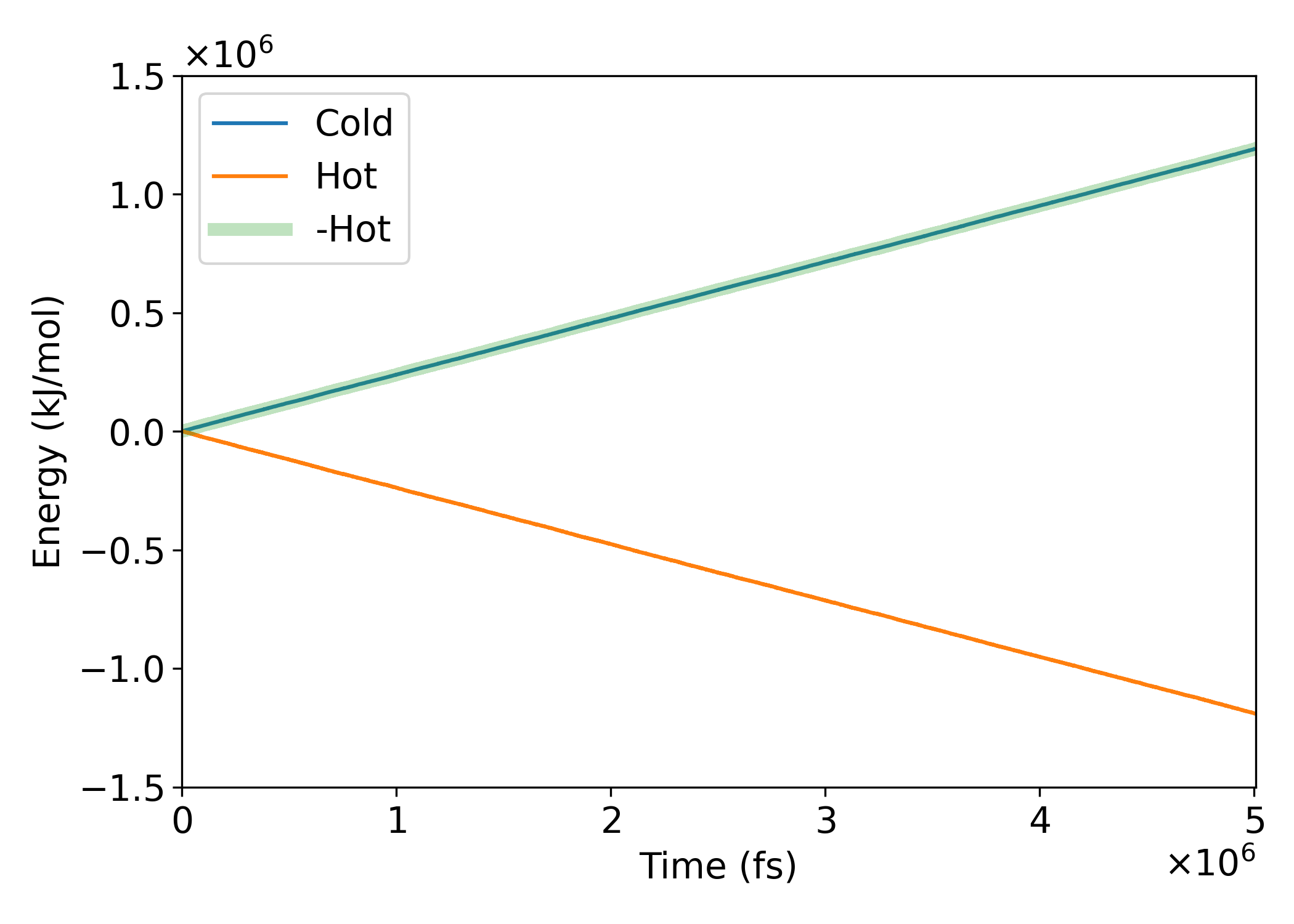}
\caption{Cumulative energy exchanged in the thermostatting regions during a 5~ns representative NEMD simulation of a 1~m LiI aqueous solution at an average temperature of 241~K and a density of 1.089~g\,cm$^{-3}$. The cold and hot reservoirs were maintained at 290~K and 190~K, respectively. The blue and orange curves denote the cumulative energy exchanged at the cold and hot thermostatting regions. To facilitate the assessment of energy conservation, the green curve represents the negative of the energy exchanged at the hot thermostat.}
\label{fig:energy_conserv}
\end{figure}

Figure~\ref{fig:energy_conserv} illustrates the energy exchanged between the hot and cold thermostatting regions during a representative NEMD simulation. The energy transfer associated with the Langevin thermostats was quantified from the work performed by the Langevin forces acting on the particles within the thermostatting slabs. These forces consist of a velocity-dependent friction term and a stochastic component governed by the fluctuation-dissipation theorem~\cite{schneider_molecular-dynamics_1978,brunger_stochastic_1984}. The close agreement between the cumulative energy exchanged at the hot and cold reservoirs demonstrates that energy is conserved within numerical accuracy throughout the simulation.

\section{Temperature Rescaling Based on Inversion Temperatures}\label{secB:rescale}

To facilitate a direct comparison between simulation and experiment, we apply a temperature rescaling based on the inversion temperature, \(T_{inv}\), defined by \(s_T(T_{inv})=0\). The inversion temperatures for both simulation and experimental datasets were determined from local linear regression fits of the Soret coefficient near the zero crossing. The simulation temperatures were then shifted by \(\Delta T = T_{inv}^{exp} - T_{inv}^{sim}\), aligning the inversion points of the two datasets. This procedure enables a direct assessment of the consistency between the simulated and experimental temperature dependence of the Soret coefficient.

\begin{figure}[H]
    \centering
    \includegraphics[width=0.7\linewidth]{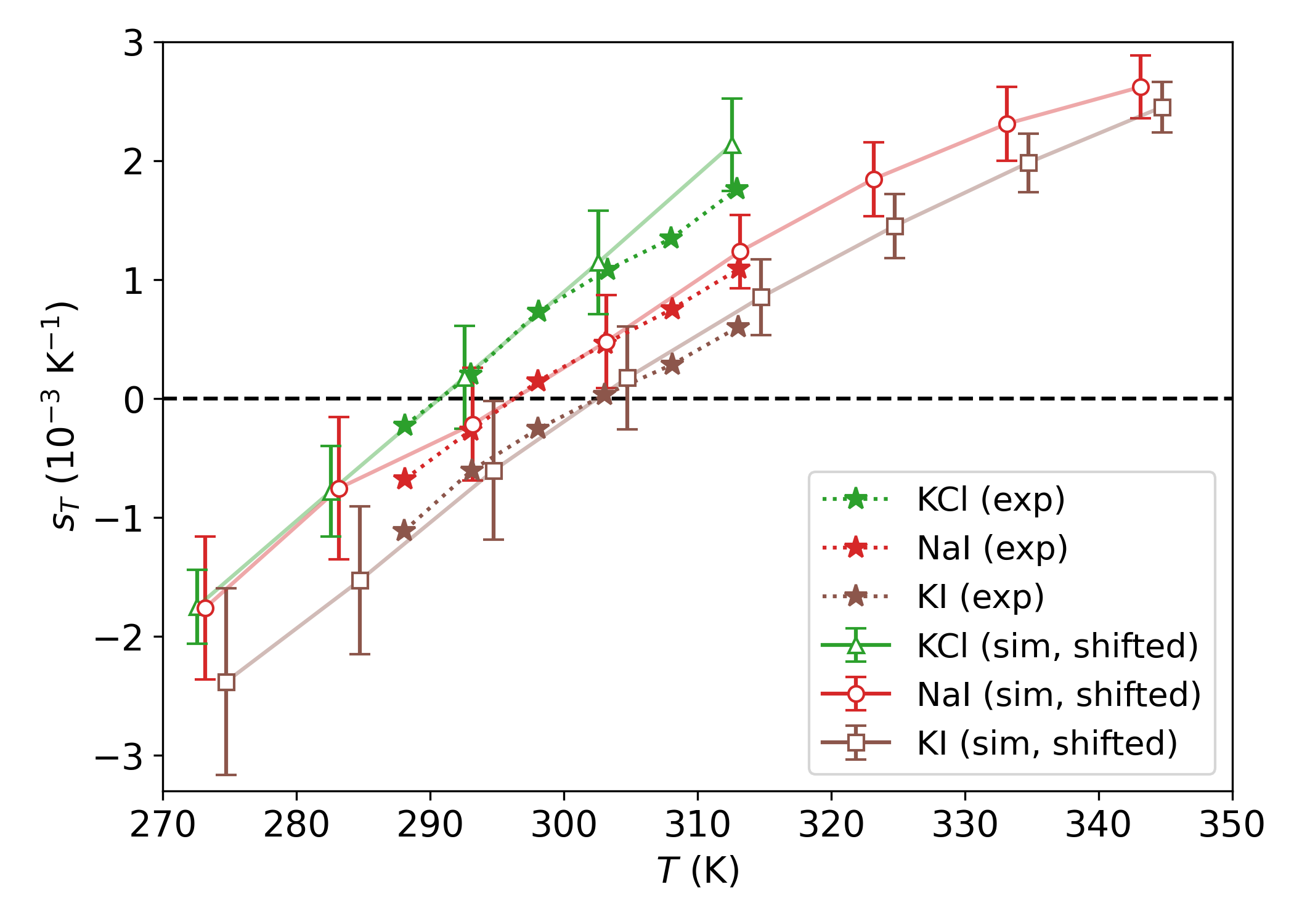}
    \caption{Soret coefficients of aqueous KCl, NaI, and KI solutions at 1~m as a function of temperature. Experimental data (stars, dotted lines) are taken from Ref.~\cite{mohanakumar_overlapping_2022} for NaI and KI, and Ref.~\cite{mohanakumar_thermodiffusion_2021} for KCl. Simulation results (symbols, solid lines) are shown after shifting the temperature to match the experimental inversion temperature, defined by $s_T(T_{\mathrm{inv}})=0$. The applied shifts are $\Delta T = 2.6$~K (KCl), $33.2$~K (NaI), and $34.8$~K (KI). These salts are the only ones in this study for which $T_{\mathrm{inv}}$ is directly accessible from experimental data. The horizontal dashed line indicates \(s_T=0\).}
    \label{fig:shifted_T}
\end{figure}

\section{Oxygen--oxygen radial distribution functions of pure water}\label{secA:OOrdf}

We show in this section the oxygen--oxygen radial distribution functions, \(g_{\mathrm{O-O}}(r)\), of pure water at 240 and 300~K, which were used as baseline data to compute the \(\Delta g_{\mathrm{O-O}}(r)\) profiles discussed in Fig~\ref{fig:O-O_rdf}.

\begin{figure}[H]
    \centering
    \includegraphics[width=0.7\linewidth]{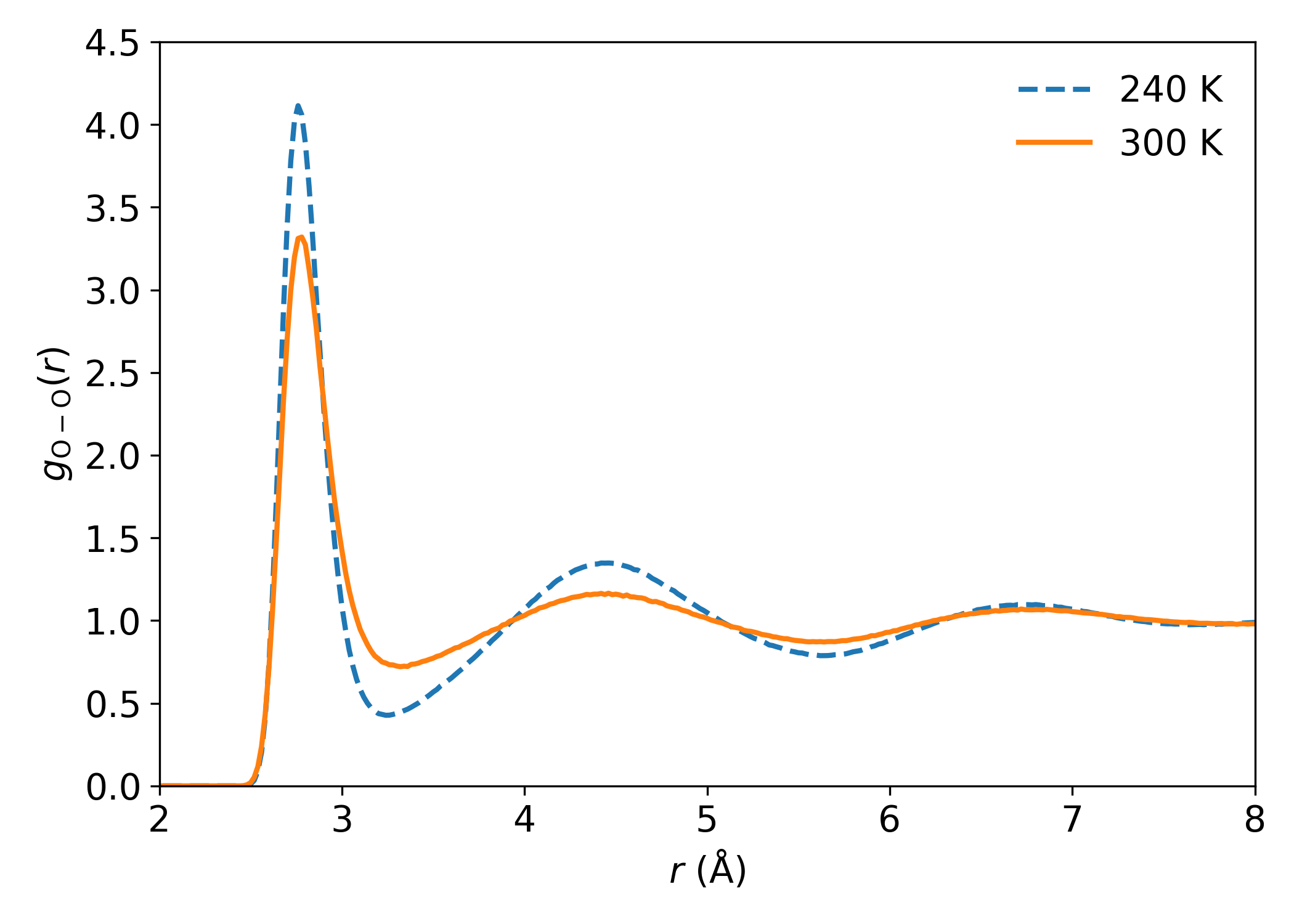}
    \caption{Oxygen--oxygen radial distribution functions, \(g_{\mathrm{O-O}}(r)\), of pure water at 240 and 300~K.}
    \label{fig:water_goo_reference}
\end{figure}

\section{NEMD Simulation Results}\label{secA:table}

The results of the NEMD simulations are reported in Table~\ref{tab:nemd}.

\sisetup{
    detect-all,
    input-symbols = (),
    table-align-text-post = false
}

\begin{longtable}{@{}lcccccccc@{}}
\caption{Summary of NEMD simulation results for aqueous KCl, LiI, NaI, and KI solutions at molalities of 1 and 4~m. The table reports the mean temperature of the simulation cell $T$, density $\rho$, pressure $P$, thermal conductivity $\lambda$, Soret coefficient $s_T$, average tetrahedral order parameter $\overline{\xi}$, and the percentage of low-density water liquid (LDL) population obtained from the tetrahedral order parameter distributions. Values correspond to time averages over steady-state simulation windows, with standard deviations across independent replicas given in parentheses.}
\label{tab:nemd}\\
\toprule
$m$ & $T$ & $\rho$ & $P$ & $\lambda$ & $s_T$ & $\overline{\xi}$ & LDL \\
{[mol\,kg$^{-1}$]} & {[K]} & {[g\,cm$^{-3}$]} & {[bar]} & {[W\,m$^{-1}$\,K$^{-1}$]} & {$[10^{-3}\,\mathrm{K}^{-1}]$} &  & {[\%]}  \\
\midrule
\endfirsthead

\multicolumn{8}{@{}l}{\tablename\ \thetable\ (continued)}\\
\toprule
$m$ & $T$ & $\rho$ & $P$ & $\lambda$ & $s_T$ & $\overline{\xi}$ & LDL \\
{[mol\,kg$^{-1}$]} & {[K]} & {[g\,cm$^{-3}$]} & {[bar]} & {[W\,m$^{-1}$\,K$^{-1}$]} & {$[10^{-3}\,\mathrm{K}^{-1}]$} &  & {[\%]}  \\
\midrule
\endhead

\bottomrule
\endlastfoot

\multicolumn{8}{c}{KCl} \\ 
\midrule
1 & 241.0 & 1.043 & 16.4(9.3) & 0.777(0.002) & -5.36(0.62) & 0.70 & 72.9 \\
1 & 300.1 & 1.042 & 8.1(0.5)  & 0.824(0.002) & 1.14(0.44)  & 0.62 & 63.4 \\
4 & 240.2 & 1.168 & 8.3(5.6)  & 0.721(0.004) & -3.85(0.28) & 0.57 & 41.6 \\
4 & 300.1 & 1.152 & 4.9(1.1)  & 0.753(0.003) & 1.38(0.27)  & 0.52 & 23.9 \\
\midrule

\multicolumn{8}{c}{LiI} \\
\midrule
1 & 241.1 & 1.089 & 14.3(5.2) & 0.754(0.004) & -3.52(0.49) & 0.71 & 74.7 \\
1 & 300.3 & 1.090 & 7.6(0.6)  & 0.802(0.003) & 0.95(0.34)  & 0.63 & 61.0 \\
4 & 240.5 & 1.350 & -15.8(3.0) & 0.649(0.004) & -2.80(0.29) & 0.57 & 45.0 \\
4 & 300.1 & 1.334 & -5.8(0.6) & 0.682(0.002) & -0.66(0.27) & 0.51 & 25.9 \\
\midrule

\multicolumn{8}{c}{NaI} \\
\midrule
1 & 241.0 & 1.112 & 13.4(8.3)  & 0.761(0.004) & -1.76(0.60) & 0.70 & 73.2 \\
1 & 300.2 & 1.108 & 6.8(0.7)   & 0.806(0.002) & 2.31(0.31)  & 0.62 & 57.7 \\
4 & 240.2 & 1.423 & -22.2(7.1)  & 0.660(0.004) & -0.74(0.38) & 0.56 & 36.2 \\
4 & 300.1 & 1.391 & -9.6(0.9)  & 0.690(0.002) & 1.10(0.21)  & 0.52 & 23.5 \\
\midrule

\multicolumn{8}{c}{KI} \\
\midrule
1 & 241.0 & 1.112 & 14.2(7.0) & 0.743(0.002) & -2.38(0.79) & 0.70 & 72.4 \\
1 & 300.3 & 1.113 & 8.7(0.8)  & 0.790(0.002) & 1.98(0.25)  & 0.62 & 61.9 \\
4 & 240.7 & 1.415 & 21.3(5.0)  & 0.613(0.004) & -2.46(0.63) & 0.56 & 38.0 \\
4 & 300.4 & 1.396 & -8.3(0.7)  & 0.651(0.003) & 1.95(0.27)  & 0.51 & 22.9 \\
\midrule
\end{longtable}

\end{appendices}

\bibliography{mybibliography.bib}
\end{document}